\documentclass{aa}  
\usepackage[varg]{txfonts}
\usepackage{natbib}
\usepackage{graphicx}
\usepackage{float}
\usepackage{comment}
\usepackage{longtable}

\begin{document}

      \title{Effect of low-mass galaxy interactions on their star formation}
   \author{Smitha Subramanian\inst{1}\fnmsep\thanks{smitha.subramanian@iiap.res.in}, 
           Chayan Mondal,$^{2}$ and 
           Venu Kalari$^{3}$
          }
   \institute{$^1$ Indian Institute of Astrophysics, Koramangala II Block, Bangalore-560034, India\\
   $^{2}$
Inter-University Centre for Astronomy and Astrophysics, Ganeshkhind, Post Bag 4, Pune 411007, India\\
$^{3}$
Gemini Observatory, NSF NOIRLab, Casilla 603, La Serena, Chile
             }

   \date{Received---- ; accepted----}

\abstract
  {According to the $\Lambda$ cold dark matter model of galaxy formation, the hierarchical assembly process is scale-free and interactions between galaxies in all mass ranges are expected. The effects of interactions between dwarf galaxies on their evolution are not well understood. In this study, we aim to understand the effect of low-mass galaxy interactions on their star formation rate (SFR).  
   We estimated the SFR of 
   22 interacting and 36 single gas-rich dwarf galaxies in the Lynx-Cancer void region using their far-ultraviolet (FUV) images from the GALEX mission. 
   We find an enhancement in SFR by a factor of  3.4$\pm$1.2 for interacting systems compared to single dwarf galaxies in the stellar mass range of 10$^{7}$ - 10$^{8}$ M$\odot$. Our results indicate that dwarf--dwarf galaxy interactions can lead to an enhancement in their SFR. These observations are similar to the predictions based on the simulations of dwarf galaxies at lower redshifts. 
   Future deeper and higher-spatial-resolution UV studies will help us to understand the effect of dwarf galaxy interactions on the spatial distribution of star forming clumps and to identify star formation in tidal tails.}

\titlerunning{Effect of dwarf-dwarf galaxy interactions on their star formation rate}
\authorrunning{Subramanian et al.}
   \maketitle
%

\section{Introduction}
One of the key drivers of galaxy evolution is the interaction between galaxies. Galaxy interactions can be broadly classified into two: mergers and fly-bys. Depending on the mass ratio of the interacting systems, they are again subclassified as minor (stellar mass ratio $<$ 1:4) and major (stellar mass
ratio $>$ 1:4) interactions. These events in the high mass regime (stellar mass $>$ 10$^{10}$ M$_{\odot}$) are relatively well studied in the local Universe, both observationally (\citealt{2000ApJ...530..660B,2010MNRAS.407.1514E,2010AJ....139.1857W,2011MNRAS.412..591P,2012MNRAS.426..549S,2013MNRAS.433L..59P,2016MNRAS.461.2589P,2016ApJS..222...16C,2016MNRAS.462.4495H,2022MNRAS.514.3294B,2022ApJ...940....4S}) and theoretically (\citealt{1972ApJ...178..623T,1989Natur.340..687H,1991ApJ...370L..65B,1994ApJ...425L..13M,1994ApJ...431L...9M,1996ApJ...464..641M,2007A&A...468...61D,2012ApJ...746..108T,2013MNRAS.430.1901H,2019MNRAS.485.1320M, 2020MNRAS.496.1124P, 2020MNRAS.494.4969P,2023MNRAS.522.5107B,2023MNRAS.519.4966B} and references therein). These studies show that the interactions between massive galaxies induce morphological changes (converting disc galaxies to spheroids), create stellar and gaseous streams around the galaxies, trigger star formation, and induce gas inflows leading to nuclear starburst and AGN activity. Similar studies of low-mass (stellar mass $<$ 10$^{10}$ M$_{\odot}$) systems are relatively few, and most have focused on individual systems ( \citealt{2012ApJ...748L..24M,2012Natur.482..192R,2015AJ....149..114P,2016ApJ...826L..27A,Privon2017}) and numerous panchromatic surveys, which studied dwarf galaxies, 
did not focus on the impact of interactions on their evolution. Observational studies are also limited due to the challenges in detecting these galaxies and the substructures around them. On the theoretical side, challenges exist in simulating large volumes to provide a realistic cosmological context while simultaneously resolving galaxies down to the dwarf regime.\\
According to the $\Lambda$ cold dark matter (CDM) model of galaxy formation, 
the hierarchical assembly process is scale-free and interactions between galaxies in all mass ranges are expected. Observations show that dwarf galaxies are often found in associations (\citealt{2006AJ....132..729T}; \citealt{2013A&A...559L..11B}) and cosmological simulations predict that subhalos are often accreted in small groups \citep{2008MNRAS.385.1365L}. This could explain the association of some of the Milky Way satellites with the plane of the orbit of the Magellanic Clouds (\citealt{2008ApJ...686L..61D}; \citealt{2015ApJ...805..130K}) and point to a scenario of Magellanic group infall onto the Milky Way. Dwarf galaxies are the dominant galaxy population at all redshifts \citep{2015A&A...575A..96G} and the majority of mergers are expected to be between them \citep{2010MNRAS.406.2267F}. Again, as low-mass galaxies have low tidal effects, it is not clear whether their interactions can induce star formation and morphological changes as observed for interacting massive galaxies. These objects are also more prone to environmental effects than massive galaxies (\citealt{2019A&A...625A.143V,2021MNRAS.506.2766H} and references therein). It is therefore essential to understand the effect of low-mass galaxy interactions on their evolution. Presently, there are dedicated ongoing and planned surveys to study dwarf galaxy assembly processes in the nearby Universe (\citealt{2015ApJ...805....2S, Higgs2016, Carlin2016, Annibali2020}). With improved cosmological simulations  now available (\citealt{Dubois2021,Martin2021}), we can compare the observed properties with predictions from simulations of dwarf galaxies.\\
\citet{2015ApJ...805....2S} found an enhancement in star formation rates (SFRs) by a factor of 2.3 in isolated paired dwarfs (with pair separation $<$ 50 kpc, mass ratio of the pair $<$ 10 and pair member masses in the range 10$^{7}$ - 10$^{9.7}$ M$_{\odot}$, with a median mass of 10$^{8.9}$ M$_{\odot}$) over isolated single dwarfs of similar stellar mass. 
For one interacting pair of dwarfs (dm1647+21, with enhanced star formation, in the sample of \citealt{2015ApJ...805....2S}), \citet{Privon2017} found that the star formation is widespread and clumpy in contrast to merging massive galaxies, where gas funnelling leads to nuclear starburst. Studies by \cite{2014MNRAS.445.1694L} and \cite{2020AJ....159..103K} found that the presence of tidal features correlates with star formation activity. \cite{2020ApJ...894...57S} explored the environmental influences on the SFR of low-mass galaxies (with stellar masses in the range 10$^{8}$ - 10$^{10}$ M$_{\odot}$, with a median mass of 10$^{9.5}$ M$_{\odot}$ at redshift $<$ 0.07) using the 
SDSS-IV/MaNGA Integral Field Unit (IFU) data. These authors found an enhancement in the SFR of pairs (with pair separation of $<$ 100 kpc, mass ratio of the pair of $<$ 4, and a line-of-sight kinematic separation of $
\le$ 100 kms$^{-1}$ ) in their inner regions, decreasing radially outwards. All these studies suggest that interactions on smaller scales have a role in triggering star formation and the spatial distribution of triggered star formation is significantly different from what is observed in interacting massive galaxies. 
\citet{Paudel2018} presented a catalogue of 177 interacting dwarfs with  stellar mass of $<$ 10$^{10}$ M$_{\odot}$ and redshifts of 
$<$ 0.02 (with a median mass of 10$^{9.1}$ 
M$_{\odot}$ 
and median redshift of 0.01). These  authors found that dwarf--dwarf interactions tend to prefer the low-density environment. 
However, \citet{Paudel2018} did not find any enhancement in the star formation of interacting dwarfs compared to the  star forming galaxies of the Local Volume. They also note that they mainly compiled their comparison sample data from the literature, and therefore their 
comparative study may not be as 
rigorous
as that of the comparative study 
provided by \cite{2015ApJ...805....2S} between interacting dwarf and non-interacting dwarf galaxies.
As described above, the median stellar mass of the sample of galaxies analysed in all these previous studies is  $\sim$ $\ge$ 10$^{9}$ M$_{\odot}$, which is the upper end of the low-mass regime. Therefore studies of more galaxies in different mass ranges are required.  
Studies of star formation in a carefully selected sample of interacting and non-interacting low-mass galaxies (in different mass ranges) in low-density environments will help us to understand the effect of dwarf--dwarf interactions on their star formation properties.\\
There are regions of low matter density in the Universe, known as voids. Surrounded by walls and filaments, voids are prominent features of the cosmic web and contain fewer galaxies. Numerical simulations as well as observations suggest that bluer dwarf galaxies with high specific SFRs (sSFRs) dominate the interior of these void regions (\citealt{Rojas2005, Liu2015}).
Though voids are low-density regions, they are not devoid of structures. There are subvoids within voids, which are surrounded by walls and filaments that are relatively high-density regions \citep{2013MNRAS.428.3409A}. The structure formation within voids is similar to the early stages of structure formation in the low-density Universe. 
Therefore, dwarf galaxies in voids provide an opportunity to study galaxy interactions and assembly processes on small scales. In this context, we aim to study the instantaneous SFR of interacting dwarf galaxies (stellar masses in the range of 10$^6$ - 10$^9$ M$_{\odot}$ with a majority of them in the mass range of 1-10 $\times$ 10$^7$ M$_{\odot}$) and make a comparison with the instantaneous SFR of the isolated single dwarf galaxies (with similar stellar masses) in the Lynx-Cancer void region using the GALEX far-ultraviolet (FUV) data in order to understand the effect of dwarf--dwarf interactions on their star formation properties. \\
The Lynx-Cancer void is located at the edge of the Local Volume at a distance of $\sim$ 18 Mpc. Its proximity allows us to study low-mass galaxies. Galaxies in this void tend to be metal deficient compared to galaxies in higher-density environments, and this particular void hosts some of the most metal-poor and gas-rich dwarf galaxies known (\citealt{Chengalur2013,2009ApJ...690.1797I,2011MNRAS.415.1188P,2016MNRAS.463..670P}). This suggests that these galaxies are relatively unevolved systems that may resemble galaxies in the early Universe. The study of such metal-poor, gas-rich, and interacting dwarf systems in this region will allow us to constrain interactions between small, high-redshift systems, which form the basis of the hierarchical galaxy assembly process.\\ 
 In the following section, we define the sample and data. In Sections 3 and 4, we present data analysis, our results and a discussion. 
 We summarise our findings in Section 5.  

\begin{table*}
\caption{Basic parameters of the sample galaxies. Parameters are taken from \citet{2011MNRAS.415.1188P},  \citet{2014AstBu..69..247P}, and \citet{2016A&A...596A..86P}.The first 22 galaxies are those that are interacting.}
\resizebox{\textwidth}{!}{%
\begin{tabular}{|c|c|c|c|c|c|c|c|c|c|}
\hline
Sl.no & Name & Ra & Dec & Distance & Holmberg  &  Stellar mass  & HI mass & Gas& Remarks \\ 
& or prefix & hh:mm:ss & dd:mm:ss & (Mpc)& radius (")  & $\times$ 10$^7$ ($M\odot$) & $\times$ 10$^7$ ($M\odot$) & fraction &  \\
\hline
\hline
\label{sample_para}
1 & SDSS & 07 23 01.420 & $+$36 21 17.100 & 16.0 & 39.2   & 2.76 &22.60 & 0.92  & pair of 2\\ 
2 & SDSS & 07 23 13.460 & $+$36 22 13.000 & 16.0 & 9.1  & 0.07 & 9.61 & 0.99 & pair of 1\\ 
3 & MCG9-13-52 & 07 46 56.360 & $+$51 17 42.800 & 10.1 & 32.3  &  1.32 &6.25 & 0.86 & pair of 4\\ 
4 & MCG9-13-56 & 07 47 32.100 & $+$51 11 29.000 & 10.0 & 32.8   & 2.46 &13.69 & 0.88 & pair of 3\\
 5 & NGC2541 & 08 14 40.180 & $+$49 03 42.100 & 12.0 & 186.7  & 95.03 & 460.82 & 0.87 & pair of 6\\ 
6 & NGC 2552 & 08 19 20.140 & $+$50 00 25.200 & 11.11 & 123.2  & 41.66 & 83.01 & 0.73 & pair of 5\\ 
7 & HS 0822+3542 & 08 25 55.430 & $+$35 32 31.900 & 13.49 & 9.7  & 0.04 & 1.46 & 0.98 & pair of 8\\ 
8 & SAO0822+3545 & 08 26 05.590 & $+$35 35 25.700 & 13.49 & 15.1  & 0.16  & 4.30 & 0.97 & pair of 7\\ 
9 & SDSS & 08 52 33.750 & $+$13 50 28.300 & 23.08 & 20.6  & 2.98 & 26.40 & 0.92 & pair of 10\\ 
10 & SDSS & 08 52 40.940 & $+$13 51 56.900 & 23.08 & 6.4  & 0.22 & - &  - & pair of 9\\ 
11 & UGC4704 & 08 59 00.280 & $+$39 12 35.700 & 11.74 & 200.5  & 6.73 & 72.86 & 0.94 & pair of 12\\ 
12 & SDSS & 08 59 46.930 & $+$39 23 05.600 & 11.63 & 19.5  & 0.96 & 1.56 & 0.69 & pair of 11\\
13 & UGC4722 & 09 00 23.540 & $+$25 36 40.600 & 27.89 & 122.8  & 22.59 & 212.9 & 0.93 & Merger \\ 
14 & KUG0934+277 & 09 37 47.650 & $+$27 33 57.700 & 25.16 & 29.6  & 7.35 & 45.42 & 0.89 & Pair\\ 
15 & UGC5272b & 09 50 19.490 & $+$31 27 22.300 & 10.27 & 15.8  & 0.19 & 2.89 & 0.95 & pair of 16\\ 
16 & UGC5272 & 09 50 22.400 & $+$31 29 16.000 & 10.3 & 86.2  & 3.41 & 46.33 & 0.95 & pair of 15\\ 
17 & UGC5540 & 10 16 21.700 & $+$37 46 48.700 & 19.16 & 81.1  & 26.89 & 46.55 & 0.70 & pair of 18\\ 
18 & HS 1013+3809 & 10 16 24.500 & $+$37 54 46.000 & 19.3 & 18.1  & 0.33 & 13.28 & 0.98 & pair of 17\\ 
19 & UGC3672 & 07 06 27.560 & $+$30 19 19.400 & 16.93 & 47.0  & 2.6 & 79.50 & 0.98 & Part of a triplet \\ 
20 & UGC3860 & 07 28 17.200 & $+$40 46 13.000 & 7.81 & 56.1  & 2.13 & 16.92 & 0.91 & Tidal features  \\ 
21 & UGC4117 & 07 57 25.980 & $+$35 56 21.000 & 14.12 & 41.6  & 3.57 & 23.73 & 0.90 & Tidal features \\ 
22 & DDO68 & 09 56 45.700 & $+$28 49 35.000 & 9.86 & 80.7   & 1.5 & 66.35 & 0.98 & Tidal features  \\ 
23 & UGC3600 & 06 55 40.000 & +39 05 42.800 & 9.3 & 74.4   & 1.91 & 11.23 & 0.89 & Isolated\\
24 & SDSS & 07 30 58.900 & $+$41 09 59.800 & 15.7 & 21.3   & 2.36 & 4.36 & 0.71 & Isolated\\ 
25 & SDSS & 07 37 28.470 & +47 24 32.800 & 10.42 & 19.6  & 0.11 &2.51 & 0.97 & Isolated\\ 
26 & UGC3966 & 07 41 26.000 & $+$40 06 44.000 & 8.64 & 52.3   & 1.22 &44.26 &0.98 & Isolated\\ 
27 & SDSS & 07 44 43.720 & +25 08 26.600 & 12.95 & 10.0   & 0.3 &1.38 & 0.86 & Isolated\\ 
28 & UGC4148 & 08 00 23.680 & $+$42 11 37.000 & 13.55 & 65.6   & 2.95 &54.15 &0.98 &Isolated\\ 
29 & NGC2500 & 08 01 53.300 & +50 44 15.400 & 10.88 & 117.0   & 126.82 &96.60 & 0.50 &Isolated\\ 
30 & MCG7-17-19 & 08 09 36.100 & $+$41 35 40.000 & 13.37 & 44.8  & 5.74  &25.31 &0.85 &Isolated\\ 31 & SDSS & 08 10 30.650 & +18 37 04.100 & 23.05 & 15.5   & 1.15 &6.27 &0.88 &Isolated\\ 
32 & SDSS & 08 12 39.530 & $+$48 36 45.400 & 11.05 & 20.0  & 0.37 &4.85 & 0.95&Isolated\\ 
33 & NGC2537 & 08 13 14.730 & $+$45 59 26.300 & 9.86 & 78.1   & 120.2 &48.44 &0.35 &Isolated\\ 
34 & IC2233 & 08 13 58.930 & $+$45 44 34.300 & 10.7 & 380.5   & 25.25 & 54.30&0.74 &Isolated\\ 
35 & UGC4426 & 08 28 28.530 & $+$41 51 22.800 & 10.3 & 69.4   & 6.18 &27.05 &0.85 &Isolated\\
36 & SDSS & 08 31 41.210 & $+$41 04 53.700 & 11.64 & 13.0   & 0.59 &0.58 &0.57 &Isolated\\ 
37 & SDSS & 08 43 37.980 & $+$40 25 47.200 & 12.05 & 12.9   & 0.41 &0.62 &0.67 &Isolated\\ 
38 & SDSS & 09 11 59.430 & $+$31 35 35.900 & 13.52 & 9.1   & 0.24 &1.81 &0.91 &Isolated\\ 
39 & IC2450 & 09 17 05.270 & $+$25 25 44.900 & 25.47 & 62.6   & 185.06 & 12.4 &  0.08 &Isolated\\ 
40 & SDSS & 09 26 09.450 & $+$33 43 04.100 & 10.63 & 47.9 & 0.13 &6.77 &0.99 & Isolated\\
41 & SDSS & 09 28 59.060 & $+$28 45 28.500 & 19.9 & 24.7   & 8.32 & 23.37& 0.79& Isolated\\ 
42 & SDSS & 09 29 51.830 & $+$11 55 35.700 & 24.29 & 31.4   & 2.76 &45.94  &0.96 &Isolated\\ 
43 & SDSS & 09 31 36.150 & $+$27 17 46.600 & 23.6 & 18.2   & 1.06 & 7.09 & 0.90 &Isolated\\ 
44 & SDSS & 09 40 03.270 & $+$44 59 31.700 & 20.71 & 10.7   & 1.6 & 1.1 & 0.7 &Isolated\\ 
45 & KISSB23 & 09 40 12.670 & $+$29 35 29.300 & 10.21 & 36.4   & 1.29 & 5.41 & 0.85 & Isolated\\ 46 & UGC5186 & 09 42 59.100 & $+$33 16 00.200 & 10.77 & 50.2  & 2.08 &2.57 & 0.62 & Isolated\\
47 & SDSS & 09 43 42.970 & $+$41 34 08.900 & 22.77 & 16.3   & 2.47 &4.89 &0.73 &Isolated\\ 
48 & SDSS & 09 44 37.110 & $+$10 00 46.300 & 22.22 & 24.3   & 3.88 &25.98 & 0.90 & Isolated\\ 49 & UGC5209 & 09 45 04.200 & $+$32 14 18.200 & 10.55 & 29.8   & 1.87 &4.02 &0.74 &Isolated\\ 
50 & SDSS & 09 47 18.350 & $+$41 38 16.400 & 22.56 & 12.9   & 0.46 &1.2 &0.78 &Isolated\\ 
51 & SDSS & 09 47 58.450 & $+$39 05 10.100 & 25.21 & 15.9  & 17.66 &104.99 & 0.89 &Isolated\\
52 & SDSS & 09 51 41.670 & $+$38 42 07.300 & 23.07 & 16.7   & 1.77 &6.02 &0.82 &Isolated\\ 
53 & SDSS & 09 54 50.600 & $+$36 20 01.900 & 10.22 & 14.8 & 0.16 &0.62 &0.84 &Isolated\\ 
54 & PC0956+4751 & 09 59 18.600 &  $+$47 36 58.400 & 18.66 & 29.0  & 1.91 &14.87 &0.91 &Isolated\\ 
55 & KUG0959+299 & 10 02 23.180 & $+$29 43 33.300 & 13.48 & 14.8   & 0.58  & -- & -- & Isolated\\ 
56 & UGC5427 & 10 04 41.050 & $+$29 21 55.200 & 9.79 & 49.9   & 5.2 & 5.98 & 0.60 &Isolated\\ 
57 & UGC 5464 & 10 08 07.700 & $+$29 32 34.400 & 16.9 & 54.8  & 9.58 &19.29 & 0.73 &Isolated\\ 
58 & SDSS & 10 10 14.960 & $+$46 17 44.100 & 18.58 & 12.9  & 0.57 & 0.49 & 0.53 &Isolated\\ 
\hline
\end{tabular}
}
\end{table*}
\section{Sample and data}
We used a catalogue of dwarf galaxies in the  Lynx-Cancer void region, provided by  \citet{2011MNRAS.415.1188P}. The catalogue consists of 79 galaxies, including 75 dwarf galaxies $(-11.9 > M_{B} > -18.0)$ and 4 sub-luminous galaxies $(-18.0
> M_{B} > -18.4)$. This catalogue is nearly complete to $M_B $<$ -14$ mag but misses some of the  faint low-surface-brightness (LSB) galaxies.  According to \cite{2011MNRAS.415.1188P}, there could be approximately 25–30 objects missing in the magnitude range of $-12.0 > M_{B}  > -13.5$ mag. The main
observational parameters of these galaxies taken from the literature are also provided in the catalogue. None of these galaxies are found to have any massive or luminous ($M_{B} < -19 $) neighbouring galaxy within a 3D distance of 2 Mpc. The sample includes nine pairs of dwarf galaxies and six single galaxies with tidal tails, suggesting their interacting nature. The authors classified these 24 galaxies (9 $\times$ 2 $+$ 6) as interacting either based on the presence of a common HI envelope around two nearby galaxies in HI maps and/or indications of perturbed morphology in the HI maps and/or optical images. The pairs have a typical projected distance separation of several tens of kiloparsecs. The isolated single dwarf galaxies (55/79) are those with no nearby dwarf galaxies and with no significant distortions indicating any past interaction.\\ 
The photometric properties of these sample galaxies are estimated and tabulated by \citet{2014AstBu..69..247P}. These authors used the SDSS images of the sample galaxies in $\it{u,g,r,i}$ bands and estimated the integrated magnitudes and colours. The B band magnitudes are also estimated from the $\it{g,r}$ magnitudes and the corresponding transformation equations. From the B-band surface brightness profiles, the authors estimated the optical and Holmberg radii (corresponding to the B-band surface brightness levels of 25 mag arcsec$^{-2}$ and 26.5 mag arcsec$^{-2}$ respectively), the effective radii, and the observed ellipticity (semi-minor axis/semi-major axis, b/a) of each galaxy. From the integrated magnitudes, \citet{2014AstBu..69..247P} also estimated the stellar mass of these galaxies using the mass--luminosity--colour relations (g band luminosity and (g-i) colour) given by \citet{2009MNRAS.400.1181Z}. We note such stellar mass estimations are valid for diverse galaxy populations, including LSB dwarf galaxies (\citealt{LSBmass2020}). As SFR varies as a function of stellar mass, comparison of SFR between the interacting and isolated galaxies in our sample should be performed for those in similar stellar mass bins. Therefore, galaxies without stellar mass estimates are removed from our further analysis. Of the 79 galaxies in the catalogue by \citet{2011MNRAS.415.1188P}, only 64 galaxies have stellar mass estimates from \citet{2014AstBu..69..247P}, as the remaining 15 galaxies are outside the SDSS footprint used to obtain multi-band photometry and hence stellar mass estimates.  Our target galaxies have a range in  stellar mass from 10$^6$ to 10$^9$ M$_{\odot}$ with the majority of them in the mass range of 1-10 $\times$ 10$^7$ M$_{\odot}$. 
We note that in the stellar mass estimates by \cite{2014AstBu..69..247P}, the galactic extinctions were not taken into account and an updated version of stellar masses by Perepelitsyna et al. (2014) is now available \footnote{https://arxiv.org/pdf/1408.0613.pdf}. 
We used the stellar mass estimates provided in the updated version.  \\ 
Our aim is to estimate and compare the instantaneous SFR of interacting and isolated dwarf galaxies in the sample. Young and massive OB-type stars emit significant amounts of radiation in the UV band. Imaging in UV will help to locate these populations and is therefore ideal for studying the star formation properties of galaxies. Of the 64 dwarf galaxies in our sample with stellar mass estimates, 58 are observed in the FUV ($\lambda_{eff}$=1538.6 {\AA}) band using the NASA GALEX mission \citep{Martin_2005}. The GALEX FUV channel imaging is at a spatial resolution of $\sim$ 4.2"  \citep{Morrissey_2007}. The archival calibrated science-ready GALEX images of sample galaxies are obtained from the MAST data archive and are used for our further analysis. If a sample galaxy has multiple GALEX FUV channel observations, then we take the image corresponding to the highest exposure time, with exposure times ranging between $\sim$ 100 s and $\sim$ 3300 s.\\ 
The basic parameters (taken from \citealt{2011MNRAS.415.1188P},  \citealt{2014AstBu..69..247P} and \citealt{2016A&A...596A..86P}) of the final sample of 58 galaxies are given in Table 1. 
The distance to each galaxy is taken from \citet{2014AstBu..69..247P}. These latter authors estimated the distance  using the relation D(Mpc) = V$_{dist}$/73(kms$^{-1}$Mpc$^{-1}$), where V$_{dist}$ is the velocity corresponding to the distance and taken from \citet{2011MNRAS.415.1188P}. 
For galaxies with reliable  distances derived with photometric methods using  Cepheids or  stars from the tip of the red giant branch (TRGB), or with the surface-brightness fluctuation (SBF) method, \citet{2011MNRAS.415.1188P} computed V$_{dist}$ = 73 (kms$^{-1}$Mpc$^{-1}$) $\times$ D(Mpc). 
For galaxies for which only redshifts were available, the contribution from the large negative peculiar velocity in the region considered was corrected while estimating V$_{dist}$. The uncertainty in the distance estimates is in the range of 0.4 -- 0.6 Mpc. 
Of the final 58  galaxies, 22 are in pairs and/or have signatures of interaction. 
The integrated HI masses of most of the sample galaxies given in \citet{2014AstBu..69..247P} and \citet{2016A&A...596A..86P} were measured  using the Nançay Radio Telescope (NRT). For some galaxies, the measurements are taken from previous literature, such as \citet{2016A&A...596A..86P}. The gas fraction given in Table 1 is the ratio of the total gas mass to the total baryonic mass, where the total gas mass is 1.33 times the HI mass (0.33 times the HI mass is taken as the Helium fraction) and the total baryonic mass is the sum of stellar mass and total gas mass \citep{2014AstBu..69..247P}. The gas fraction of these galaxies suggests that they are gas-rich, with median values of 0.92 and 0.88 for interacting and isolated systems, respectively. \\
\begin{figure*}
\includegraphics[width=\columnwidth]{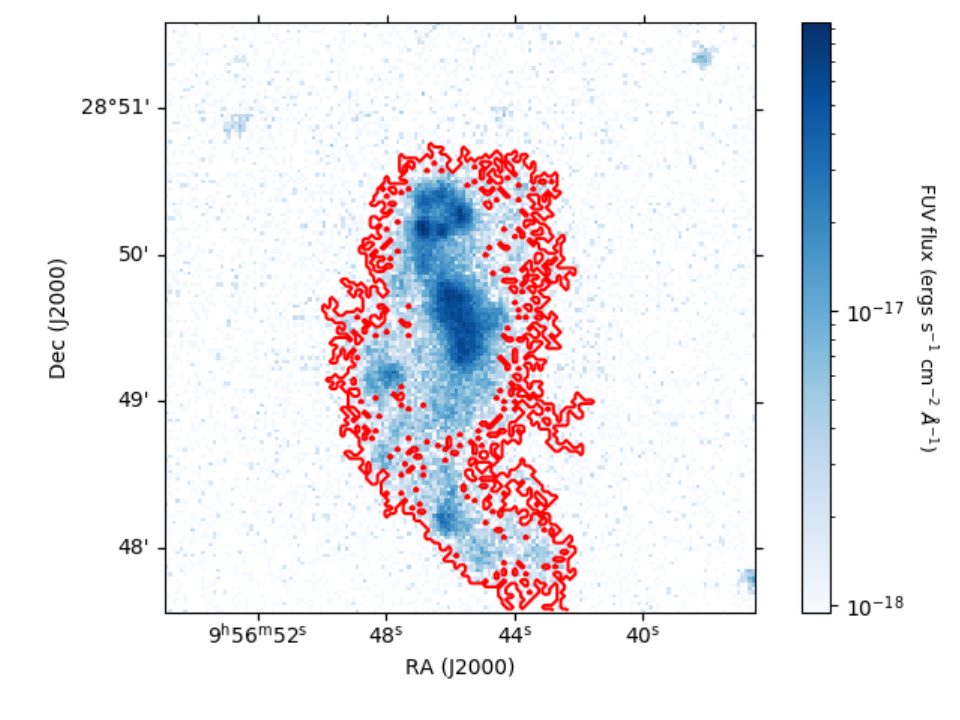}
\includegraphics[width=\columnwidth]{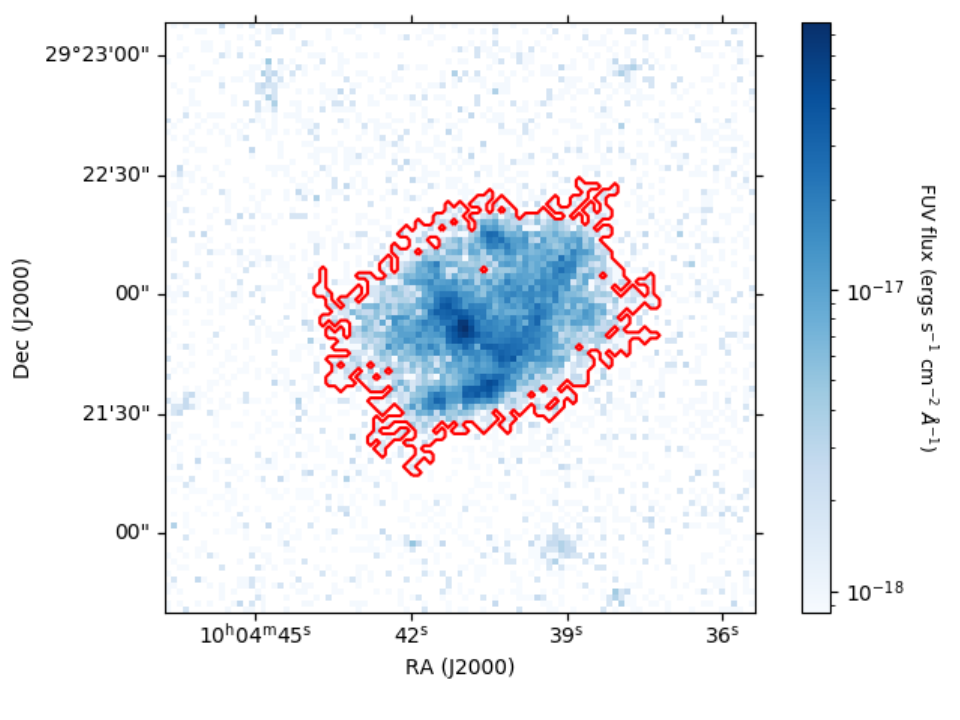}
\includegraphics[width=\columnwidth]{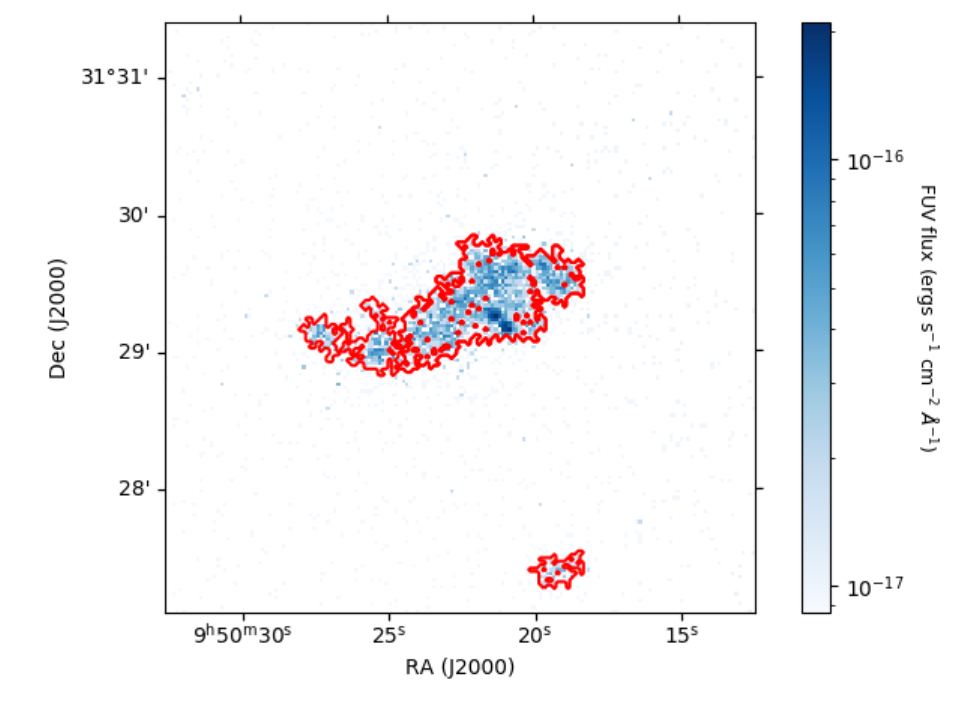}
\includegraphics[width=\columnwidth]{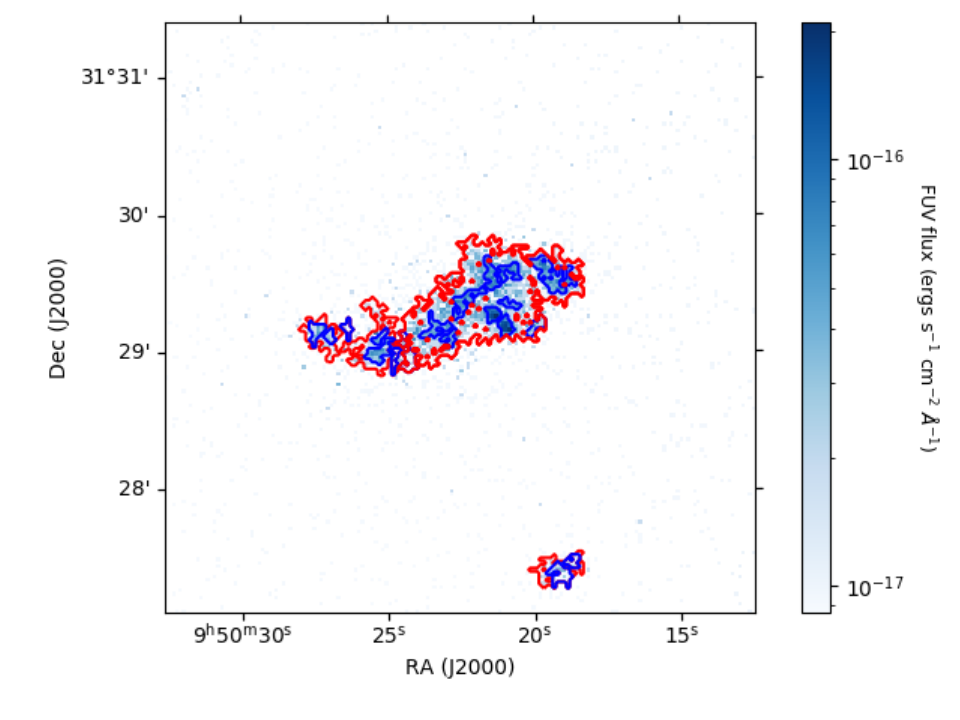}\\
   \caption{Astrodendro-identified largest structures corresponding to some of the target galaxies in the GALEX FUV images. The top-left, top-right, and bottom-left panels show the  largest structures (in red) corresponding to the galaxies DDO 68 (a galaxy classified as interacting based on the presence of tidal features), UGC 5427 (an isolated single dwarf), and the UGC 5272/5272b interacting system, respectively. In the bottom-left panel, the UGC 5272 is in the centre and UGC 5272b is $\sim$ 2 arcmin towards the south. The bottom-right panel shows the subclumps (in dark blue) identified inside the UGC 5272/5272b system.}
\end{figure*} 
\section{Analysis}
As most of the FUV photons that emerge from a galaxy originate from the atmospheres of young stars ($\sim$ 100 Myr), the current SFR of a galaxy is  proportional to the FUV luminosity emitted by the young stars. 
Under the assumption that the SFR is approximately
constant over the past 100 Myr, the observed FUV flux from star forming galaxies is a direct tracer of current star formation \citep{1998ApJ...498..541K}. We used the following equation (Equation 1), as given in \cite{2011ApJ...737...67M} for the GALEX FUV band, to estimate the current SFR of our sample of galaxies. \\\\
$SFR_{FUV}[M{\odot}/yr] = 4.42 \times 10^{-44}$   $L_{FUV}[erg/sec]$  ...... (1).\\\\
\cite{2011ApJ...737...67M} derived this formula from Starburst99 \citep{1999ApJS..123....3L},  assuming solar metallicity, and a Kroupa \citep{2001MNRAS.322..231K} IMF, with a slope of $-1.3$ and $-2.3$ for stellar masses in the ranges of $0.1 - 0.5$ $M\odot$ and $0.5 - 100$  $M\odot,$ respectively. We note that the calibration constant used in the relation can vary by up to a factor of $\sim$ 1.5 with stellar metallicity (at lower metallicities the UV luminosity increases;   
\citealt{2005A&A...443L..19B}). We do not have metallicity information for all of the galaxies in our  sample. The gas-phase metallicity values reported by \citet{2014AstBu..69..247P} for some of them are lower than solar, but are approximately similar to each other in the same stellar mass range. Therefore, our final results based on the comparison of the SFR between the interacting and isolated galaxies in the same stellar mass range as our sample is not expected to be significantly affected by the choice of calibration constant. \\
\begin{table*}[]
\centering
\caption{Derived FUV luminosity and SFR of sample galaxies. The number and properties of the star forming clumps identified in each galaxy are also given, with their median properties given in parentheses.}
\label{table2}
\begin{tabular}{|c|c|c|c|c|c|c|}
\hline
Sl.no & L$_{FUV}$ & SFR &  No of   & Size of & SFRD of clumps  \\ 
& $\times$ 10$^{39}$ erg s$^{-1}$ & $\times$ 10$^{-4}$ (M$_{\odot}$yr$^{-1}$) &clumps&clumps (pc)&($\times$ 10$^{-4}$ (M$_{\odot}$yr$^{-1}$kpc$^{-2}$))\\ \hline
1 & 18.76 $\pm$ 0.20 &  8.29 $\pm$ 0.09 & 4 & 227.35 -- 359.47 ( 284.47 ) & 6.36 -- 8.85 ( 7.41 ) \\
2 & 4.15 $\pm$ 0.07 &  1.83 $\pm$ 0.03 & 1 & 334.65  & 5.22  \\
3 & 12.02 $\pm$ 0.20 &  5.31 $\pm$ 0.09 & 5 & 143.51 -- 198.69 ( 175.77 ) & 9.74 -- 12.55 ( 10.57 ) \\
4 & 141.50 $\pm$ 1.08 &  62.54 $\pm$ 0.48 & 4 & 147.89 -- 246.11 ( 188.61 ) & 12.93 -- 98.76 ( 65.53 ) \\
5 & 1378.40 $\pm$ 0.62 &  609.25 $\pm$ 0.27 & 198 & 155.65 -- 767.30 ( 225.57 ) & 0.78 -- 34.68 ( 2.26 ) \\
6 & 518.54 $\pm$ 0.49 &  229.19 $\pm$ 0.22 & 89 & 144.11 -- 407.61 ( 203.80 ) & 1.00 -- 23.55 ( 3.47 ) \\
7 & 19.38 $\pm$ 0.06 &  8.57 $\pm$ 0.03 & 1 & 479.21  & 11.88  \\
8 & 13.78 $\pm$ 0.05 &  6.09 $\pm$ 0.02 & 1 & 727.81  & 3.66  \\
9  & 30.27 $\pm$ 0.18 &  13.38 $\pm$ 0.08 & 4 & 366.66 -- 912.97 ( 401.50 ) & 1.44 -- 3.52 ( 1.93 ) \\
10 & 3.44 $\pm$ 0.04 &  1.52 $\pm$ 0.02 & 1 & 491.92  & 2.00  \\
11 & 135.30 $\pm$ 0.26 &  59.80 $\pm$ 0.12 & 21 & 152.28 -- 502.76 ( 198.55 ) & 0.59 -- 15.70 ( 3.05 ) \\
12 & 8.68 $\pm$ 0.04 &  3.83 $\pm$ 0.02 & 2 & 297.92 -- 518.21 ( 408.06 ) & 0.98 -- 3.91 ( 2.45 ) \\
13 & 840.09 $\pm$ 5.12 &  371.32 $\pm$ 2.26 & 5 & 594.44 -- 816.99 ( 667.07 ) & 19.57 -- 37.63 ( 22.33 ) \\
14  & 76.52 $\pm$ 1.06 &  33.82 $\pm$ 0.47 & 3 & 326.36 -- 765.37 ( 565.27 ) & 6.47 -- 10.25 ( 10.04 ) \\
15  & 7.72 $\pm$ 0.11 &  3.41 $\pm$ 0.05 & 2 & 151.89 -- 230.73 ( 191.31 ) & 7.64 -- 9.20 ( 8.42 ) \\
16  & 226.60 $\pm$ 1.22 &  100.16 $\pm$ 0.54 & 21 & 133.60 -- 307.58 ( 174.20 ) & 6.26 -- 55.43 ( 18.56 ) \\
17  & 199.73 $\pm$ 2.86 &  88.28 $\pm$ 1.26 & 5 & 248.53 -- 745.59 ( 324.04 ) & 15.97 -- 41.22 ( 23.04 ) \\
18  & 221.80 $\pm$ 2.92 &  98.04 $\pm$ 1.29 & 1 & 755.19  & 54.74 \\
19  & 107.78 $\pm$ 0.79 &  47.64 $\pm$ 0.35 & 2 & 568.43 -- 839.10 ( 703.76 ) & 10.90 -- 16.07 ( 13.49 ) \\
20  & 40.08 $\pm$ 0.06 &  17.72 $\pm$ 0.03 & 18 & 101.31 -- 256.28 ( 143.27 ) & 1.64 -- 10.20 ( 4.99 ) \\
21  & 100.91 $\pm$ 1.24 &  44.60 $\pm$ 0.55 & 9 & 183.15 -- 474.08 ( 216.71 ) & 9.73 -- 21.92 ( 13.66 ) \\
22  & 203.75 $\pm$ 0.27 &  90.06 $\pm$ 0.12 & 42 & 127.90 -- 561.87 ( 171.59 ) & 0.47 -- 30.88 ( 1.69 ) \\
23  & 23.55 $\pm$ 0.13 &  10.41 $\pm$ 0.06 & 12 & 126.52 -- 497.38 ( 147.66 ) & 3.37 -- 6.53 ( 4.89 ) \\
24  & 40.18 $\pm$ 0.18 &  17.76 $\pm$ 0.08 & 1 & 737.08  & 10.41 \\
25  & 3.53 $\pm$ 0.05 &  1.56 $\pm$ 0.02 & 1 & 237.97  & 8.77  \\
26  & 64.99 $\pm$ 0.07 &  28.73 $\pm$ 0.03 & 26 & 117.54 -- 506.19 ( 166.23 ) & 1.14 -- 8.22 ( 2.67 ) \\
27  & 5.67 $\pm$ 0.10 &  2.51 $\pm$ 0.04 & 1 & 314.26  & 8.09  \\
28  & 84.73 $\pm$ 0.16 &  37.45 $\pm$ 0.07 & 8 & 200.40 -- 707.42 ( 294.06 ) & 2.18 -- 8.40 ( 3.89 ) \\
29  & 1009.28 $\pm$ 0.54 &  446.10 $\pm$ 0.24 & 94 & 141.13 -- 518.53 ( 194.53 ) & 0.90 -- 48.23 ( 4.67 ) \\
30  & 113.45 $\pm$ 0.81 &  50.15 $\pm$ 0.36 & 4 & 189.98 -- 567.29 ( 266.09 ) & 6.17 -- 31.65 ( 20.68 ) \\
31  & 13.45 $\pm$ 0.06 &  5.94 $\pm$ 0.03 & 1 & 940.74  & 2.14  \\
32  & 10.17 $\pm$ 0.02 &  4.50 $\pm$ 0.01 & 1 & 623.12  & 3.69  \\
33  & 552.33 $\pm$ 0.30 &  244.13 $\pm$ 0.13 & 30 & 127.90 -- 299.94 ( 171.59 ) & 1.26 -- 151.13 ( 2.09 ) \\
34  & 410.49 $\pm$ 0.40 &  181.43 $\pm$ 0.18 & 28 & 138.79 -- 496.56 ( 180.65 ) & 0.98 -- 39.88 ( 11.82 ) \\
35  & 31.71 $\pm$ 0.07 &  14.02 $\pm$ 0.03 & 35 & 133.60 -- 387.22 ( 202.62 ) & 0.86 -- 5.07 ( 1.49 ) \\
36  & 4.80 $\pm$ 0.07 &  2.12 $\pm$ 0.03 & 1 & 301.97  & 7.41  \\
37  & 8.56 $\pm$ 0.18 &  3.78 $\pm$ 0.08 & 1 & 283.94  & 14.95  \\
38  & 13.18 $\pm$ 0.03 &  5.83 $\pm$ 0.01 & 1 & 594.71  & 5.25  \\
39  & 281.19 $\pm$ 2.67 &  124.29 $\pm$ 1.18 & 1 & 1325.63  & 22.52  \\
40  & 13.16 $\pm$ 0.23 &  5.82 $\pm$ 0.10 & 6 & 157.21 -- 242.77 ( 182.38 ) & 6.08 -- 9.89 ( 7.58 ) \\
41  & 29.74 $\pm$ 0.47 &  13.14 $\pm$ 0.21 & 1 & 642.73  & 10.13  \\
42  & 40.32 $\pm$ 0.67 &  17.82 $\pm$ 0.30 & 4 & 330.45 -- 668.37 ( 404.67 ) & 4.08 -- 6.67 ( 4.90 ) \\
43  & 15.11 $\pm$ 0.32 &  6.68 $\pm$ 0.14 & 2 & 349.03 -- 421.96 ( 385.50 ) & 4.90 -- 7.70 ( 6.30 ) \\
44  & 6.30 $\pm$ 0.25 &  2.78 $\pm$ 0.11 & 1 & 306.29  & 9.45 \\
45  & 43.88 $\pm$ 0.40 &  19.39 $\pm$ 0.18 & 1 & 539.59  & 21.21  \\
46  & 12.18 $\pm$ 0.28 &  5.38 $\pm$ 0.12 & 1 & 366.96  & 12.73  \\
47  & 21.30 $\pm$ 0.64 &  9.42 $\pm$ 0.28 & 2 & 336.76 -- 407.12 ( 371.94 ) & 9.29 -- 10.74 ( 10.01 ) \\
48  & 51.83 $\pm$ 0.14 &  22.91 $\pm$ 0.06 & 8 & 302.29 -- 455.72 ( 369.84 ) & 1.40 -- 4.66 ( 3.04 ) \\
49  & 10.71 $\pm$ 0.15 &  4.74 $\pm$ 0.07 & 3 & 173.10 -- 348.89 ( 202.98 ) & 5.25 -- 7.74 ( 6.25 ) \\
50  & 23.23 $\pm$ 0.67 &  10.27 $\pm$ 0.30 & 1 & 471.85  & 14.69  \\
51  & 6.53 $\pm$ 0.30 &  2.89 $\pm$ 0.13 & 1 & 342.97  & 7.82  \\
52  & 8.64 $\pm$ 0.34 &  3.82 $\pm$ 0.15 & 1 & 313.85  & 12.34  \\
53  & 2.11 $\pm$ 0.08 &  0.93 $\pm$ 0.03 & 1 & 192.11  & 8.07  \\
54  & 16.53 $\pm$ 0.46 &  7.31 $\pm$ 0.20 & 3 & 242.04 -- 342.30 ( 242.04 ) & 7.27 -- 11.85 ( 8.75 ) \\
55  & 10.48 $\pm$ 0.17 &  4.63 $\pm$ 0.08 & 1 & 435.38  & 7.78  \\
56  & 68.51 $\pm$ 0.13 &  30.28 $\pm$ 0.06 & 12 & 133.19 -- 295.09 ( 160.63 ) & 0.53 -- 17.64 ( 8.68 ) \\
57  & 30.58 $\pm$ 0.59 &  13.52 $\pm$ 0.26 & 5 & 229.91 -- 373.31 ( 285.82 ) & 8.64 -- 12.22 ( 9.58 ) \\
58  & 16.15 $\pm$ 0.29 &  7.14 $\pm$ 0.13 & 1 & 488.00  & 9.55  \\
\hline
\end{tabular}
\end{table*}
\begin{figure*}
\includegraphics[width=\columnwidth]{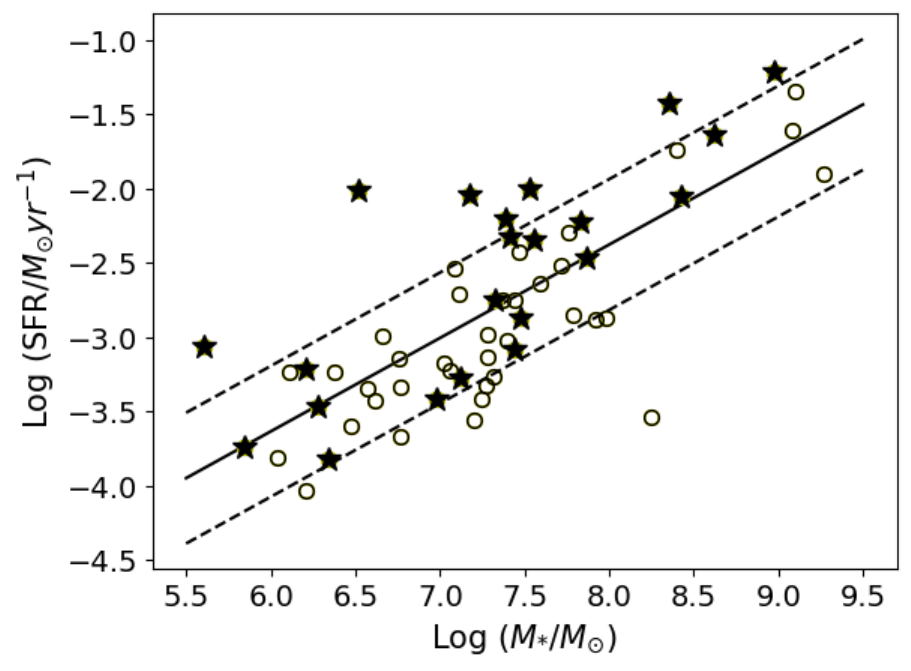}
\includegraphics[width=\columnwidth]{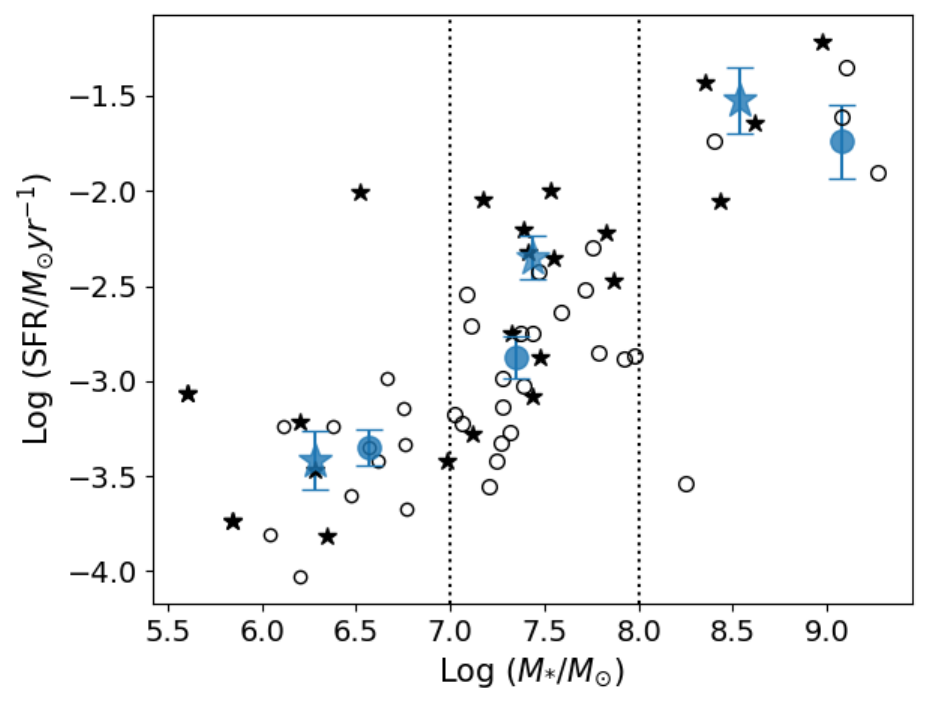}\\
   \caption{Current instantaneous SFR vs M$_{*}$ for our sample galaxies. The black stars and open circles in both panels represent the interacting and isolated dwarf galaxies, respectively. Left panel: Solid and dotted lines represent the best-fit line and the $\pm$ rms lines, respectively. Right panel: Dotted lines show the mass bin of 10$^{7-8}$ M$_{\odot}$. The blue solid star and closed circle correspond to the median log(SFR) values of the sample in different mass bins.}
   \label{img001}
\end{figure*} 
We further analysed the calibrated GALEX images containing our sample galaxies in order to obtain the integrated FUV flux and $L_{FUV}$. From the 1$^\circ$.5 field of the GALEX tile, the region containing the sample galaxy is extracted. The extracted image has the sample galaxy at the centre and has dimensions equal to three times the Holmberg radius (corresponding to the B-band surface brightness level of 26.5 mag arcsec$^{-2}$) measured by \citet{2014AstBu..69..247P} from optical images. \\ 
We identified the largest structure corresponding to the sample galaxy using the \textit{astrodendro}\footnote{https://dendrograms.readthedocs.io/en/stable/} Python package. Astrodendro identifies structures in an intensity map for a given value of threshold flux, and a minimum number of pixels. We used three times the median value of the sky background as the threshold flux. The sky background is estimated using the sky background image corresponding to the target field provided in the GALEX MAST archive. The minimum number of pixels is taken as ten, which means that the structures that cover less than 10 pixels are not considered. This value is chosen because the area covered by 10 pixels is equivalent to a circle of $\sim$ 1.8 pixels in  radius, and therefore the size of the identified  structure ($\sim$ 3.6 pixels) is comparable to or larger than the PSF of the observed field. The output of astrodendro provides the position, area, and flux of all the identified structures in a field. The largest structure (out of all the identified structures in a galaxy, the one with the maximum value for the area) corresponds to the entire galaxy and is identified. The top and bottom-left panels of Fig.1 show the astrodendro-identified largest structures (in red) corresponding to some of the target galaxies in the GALEX FUV images. The top-left panel shows the galaxy DDO68, which is classified as interacting based on the irregular morphology identified in the HI maps as well as optical images (\citealt{2008MNRAS.391..881E, 2016ApJ...826L..27A}). The top-right panel shows an isolated galaxy, UGC 5427. The bottom-left panel displays an interacting system, UGC 5272 and UGC 5272b, which has a HI bridge connecting the two galaxies \citep{2002A&A...390..829S}. The UGC 5272 is in the centre and UGC 5272b is $\sim$ 2 arcmin towards the south. The bottom-right panel shows the subclumps (in dark blue) identified inside the UGC 5272--UGC 5272b system.\\ The flux of the largest structure, provided as an output by astrodendro, is taken as the total flux of the galaxy.
To correct for the background, we used the sky background image corresponding to the target field taken from the GALEX MAST archive. The background corresponding to the galaxy is estimated by multiplying the exact area of the structure (provided by the astrodendro package) by the median background value and subtracted from the total flux. 
The measured integrated FUV flux is corrected for the Galactic foreground extinction using the values provided by \citet{1998ApJ...500..525S} and applying the calibration given by \citet{2011SchlaflyFinkbeiner}  (R$_{FUV}$ is taken as 8.06 from \citealt{2011Ap&SS.335...51B} ). The extinction-corrected integrated UV flux is converted to $L_{FUV}$ using the distance values given in Table 1. Using the estimated $L_{FUV}$ and the SFR formula given in Equation 1, we estimated the SFR of each galaxy. The estimated total luminosity and SFR of each galaxy are tabulated in Table 2. We did not correct for the internal extinction and also the R$_{FUV}$ values for low-mass galaxies can be different (generally seems to be up to $\sim$ 1.5 times higher, \citealt{2011Ap&SS.335...51B}) from the assumed value based on the Milky Way. The estimated FUV flux or luminosity is therefore a lower limit, as is the derived SFR. 
Using astrodendro, we also identified the star forming clumps (the smallest structure appearing in the GALEX image, which cannot be split further) in these galaxies. The number of clumps identified, their sizes, and their star formation rate density (SFRD) are listed in Table 2.  As the galaxies of our sample are intrinsically smaller in size, and also given the poor spatial resolution of GALEX, there are not many clumps identified in most of the sample galaxies. The structure corresponding to the entire galaxy is the only structure identified in most of the galaxies and their sizes are comparable to the optical size of these galaxies, given in Table 2 of \cite{2014AstBu..69..247P}. The clumps identified in GALEX images can actually be composed of multiple smaller clumps, which we were not able to resolve due to the limited angular resolution. Therefore, the size of individual star-forming clumps would be smaller than the values noted in Table 2. Deeper and higher-spatial-resolution UV observations (e.g. using the Ultraviolet Imaging Telescope, UVIT, on board AstroSat) are required in order to identify smaller star forming clumps, study their spatial distribution, and carry out any meaningful analysis of the SFRD values of the star forming clumps. 
\section{Results and Discussion}
To understand the effect of interactions on the SFR, we analysed the correlation between the derived instantaneous SFR and the stellar mass of the galaxies in our sample. To this end, we plotted the derived instantaneous SFR of our sample against stellar mass (left panel of Fig. 2). The stars and circles represent the interacting and isolated set of galaxies, respectively. The errors in the derived SFR are of the order of 10$^{-2}$ dex and the error bars are shorter than the size of the points and are therefore not shown. The figure shows that the SFR increases as stellar mass increases, for both the interacting and isolated galaxies. When we consider the location of the interacting systems in galaxy scaling relations to check for correlations or offsets, it is essential to understand whether the parameters are estimated for the entire system or  for individual components. 
As described in Sect. 2, there are 22 interacting galaxies in our sample and these are in different stages of interaction. There are 16 (8x2) galaxies are in pairs where both galaxies are well separated in optical/UV images but show a common envelope around them in HI and/or are at similar distances with perturbed morphology, indicating that they are pair. The current SFR of each galaxy in these eight pairs (8x2 = 16 galaxies) is calculated separately and these values are plotted against their respective stellar mass. Six single galaxies that show tidal tails and a perturbed morphology in the HI maps and/or optical images are also considered as interacting galaxies in our study. The perturbed morphology of these six single galaxies, which are classified as  interacting systems, could be the effect of a past fly-by event, with tidal features, or may indicate that the galaxy is a merger remnant with multiple smaller systems together. For these galaxies, the SFR is calculated based on the FUV flux of the entire system. Similarly, the stellar mass of these galaxies taken from the study of \cite{2014AstBu..69..247P} is calculated using the integrated colour and luminosity of the entire system from SDSS images. Specifically, these latter authors used the g-band luminosity and (g-i) colour relations provided by \cite{2009MNRAS.400.1181Z}, because this combination yields the most robust estimates in the optical part of the spectrum and the mass--g-band luminosity ratio takes into account the complex star formation history observed in low-mass galaxies including the recent episodes of star formation. Considering all these points, we expect these systems to follow the 
expected correlation between the current SFR (estimated from FUV luminosity) and stellar mass (estimated using the g-band luminosity and (g-i) colour relation). Any deviation from the expected SFR corresponding to a stellar mass could be due to a real increase or decrease in star formation. This assumption is valid even when the current SFR and the stellar estimates are dominated by that of the main galaxy of the system, and some outer regions corresponding to tidal tails or accreted systems are not considered (because of their low surface brightness).  However, this assumption will not be valid if the current SFR is concentrated only in some regions of the galaxies (which might be the tidal tails or accreted systems) and we compare it against the total stellar mass of the system, which might be dominated by the main galaxy. Based on the analysis presented here, we see that the star formation identified in all the  galaxies of our sample (including these six galaxies classified as interacting based on their perturbed morphology) is widespread over the entire system and is not concentrated to any specific regions. We therefore expect these galaxies to follow the stellar mass--SFR scaling relation and any deviation from this relation can be considered as real. We therefore performed a linear fit to the entire sample. A positive correlation between the log(SFR) and log(stellar mass) is found, with a slope of 0.62 and an intercept of $-7.4$. The best-fit line is shown as a solid black line in the figure. The rms scatter in log(SFR) is 0.44 dex and the dotted lines correspond to   $+/-$ $0.44$ dex to the best-fit log(SFR). Most of the points are within the $+/-$ rms scatter. However, the majority of the galaxies in the interacting sample are above the best-fit line (with six galaxies above the $+$ rms scatter line) and the majority of the galaxies in the isolated sample are below the best-fit line. \\

There is an insufficient number of interacting and isolated galaxies in the entire mass range to separately fit the two sets. Most of the galaxies in our sample (31/58, with 11 interacting and 20 isolated galaxies) are in the mass range of 10$^7$ - 10$^8$ M$\odot$. We selected this mass bin (as shown by two vertical dotted lines in the right panel of Fig. 2) and estimated the median log(SFR) of the interacting and isolated sample. The median SFR is 3.4$\pm$1.2 times higher for the interacting sample than that of the isolated sample in this mass bin.  The median values are shown as a transparent purple star and circle for the interacting and isolated samples, respectively. The remaining mass bins are not equally populated with the two sets. We therefore estimated the median log(SFR) of the interacting and isolated sample with stellar mass $<$ 10$^7$ M$\odot$ and $>$ 10$^8$ M$\odot$ and shown in the right panel of Fig. 2. 
The ratio of the median SFR of the interacting sample to that of the isolated sample in the mass bins with stellar mass $<$ 10$^7$ M$\odot$ and stellar mass $>$ 10$^8$ M$\odot$ is $\sim$ 0.85 and 1.65, respectively. 
Both these mass bins contain fewer galaxies compared to the middle mass bin. The galaxies of the sample in the lower mass bin (stellar mass $<$ 10$^7$ M$\odot$) have $M_B > -13.5$ mag and therefore this mass bin could be significantly affected by the incompleteness of the initial sample as described in Section 2. There is a slight enhancement of SFR for the interacting sample compared to the isolated sample in the higher mass bin (stellar mass $>$ 10$^8$ M$\odot$). We note that the stage and nature of the interaction (major or minor merger, non-merger interactions, and the projected distance between the pair) can also affect the SFR. However, the number of galaxies in our current sample is not sufficient to make a statistical and meaningful comparison between the  subclasses based on these parameters. \\ 

Most of the previous observational studies of low-mass galaxies exploring the effect of interactions on their evolution were focused on individual systems. The first systematic study to understand environmental effects on star formation using a large sample of dwarf galaxy pairs 
was performed by \cite{2015ApJ...805....2S}. Their sample covered a mass range of 10$^7$ - 5 $\times$ 10$^9$ M$\odot$ and a redshift range of 0.005 $<$ z $<$ 0.07.  
Based on the presence or absence of a massive galaxy (stellar mass $>$ 5 $\times$ 10$^9$ M$\odot$) at a distance $<$ 1.5 Mpc, these pairs were again classified as non-isolated or isolated pairs. The median mass of the sample was 10$^{8.9}$ M$\odot$. More than 90\% of the pairs in the sample had a mass ratio of $<$ 5. These latter authors observed an enhancement in the SFR of the dwarf galaxies in pairs by a factor of 2.3 ($\pm$ 0.7) compared to that of the isolated single dwarfs (matched in redshift and stellar mass), for pair separations of $<$ 50 kpc. Such an enhancement was observed for both isolated and non-isolated pairs. This suggests that close encounters between dwarf galaxies do enhance their SFR, irrespective of the presence or absence of a massive neighbour. \cite{2015ApJ...805....2S} also found that the enhancement decreases with increasing pair separation and was observed out to pair separations as far as 100 kpc for isolated dwarf pairs. 
The enhancement in SFR for dwarf pairs, by a factor of 2.3 ($\pm$ 0.7),  observed by these latter authors is comparable to the enhancement factor of 3.4$\pm$1.2 observed in the present study. We note that the sample in our study is smaller and cannot therefore be subclassified based on pair separation distance and mass ratio as in the study by \cite{2015ApJ...805....2S}. Also, the median stellar mass of our sample is $\sim$ 2.3  $\times$ 10$^{7}$ M$\odot$, whereas that of the sample in the study of \cite{2015ApJ...805....2S} is 10$^{8.9}$ M$\odot$. \\

\cite{2015MNRAS.454.1742K} studied a sample of
approximately 1500 of the nearest galaxies (with stellar masses in the range 10$^{8.0 - 11.0}$ M$\odot$) ---all within a distance of $\sim$ 45 Mpc--- to investigate the influence of interactions on star formation. This representative sample of nearby galaxies includes many low-mass galaxies (stellar mass $<$ 10$^9$ M$\odot$, but with most of them in the range 10$^{8.0 - 9.0}$ M$\odot$. These authors found that both SFR and sSFR  
(which is SFR normalised by the stellar mass of the galaxy) 
are enhanced in interacting galaxies. The increase is moderate, reaching a maximum of a
factor of 1.9 for the highest degree of interaction (mergers). A recent study by \cite{2020ApJ...894...57S} explored the environmental influence on star formation in low-mass galaxies using the SDSS-IV/MaNGA spatially and spectroscopically resolved data of 386 low-mass galaxies with stellar mass in the range 10$^{8-10}$ M$\odot$ (with median stellar mass of $\sim$ 10$^9.5$ M$\odot$) and at redshifts of 0.01 $<$ z $<$ 0.07. These authors found that star formation activities in low-mass galaxies are affected by their environment and found an enhancement in their SFR. For the pair candidates with mass ratios of between 0.25 and 4 and at projected distances of $<$ 100 kpc, they found an enhancement in SFR by a factor of 1.75 $\pm$ 0.96 in the inner regions, with this enhancement factor decreasing outwards. Though the properties of the sample (such as stellar mass and distance) in these previous studies are different from those presented in our study, the observed factors of enhancement in the SFR of interacting dwarfs compared to isolated dwarfs in these studies are comparable to the values we find for our sample.\\ 

Recently,  \cite{Martin2021}  used simulations of low-mass galaxies to investigate the effect of mergers and interactions on their star formation, and their evolution up to a redshift of 0.5. These authors found that the mergers drive a moderate enhancement in star formation (3 - 4 times at z = 1) and non-merger interactions drive a smaller enhancement in star formation ($\sim$ 2 times). However, non-merger interactions are numerous compared to major and minor mergers and therefore contribute to the stellar mass growth of dwarf galaxies. Figure 9  of this latter publication shows the average displacement of galaxies in their simulation from the best-fit star forming main sequence in the log(SFR) versus stellar mass plot. This plot shows that the galaxies in the mass range 10$^{7.5 - 9}$ M$\odot$ have a displacement of 0.45 - 0.85 dex at the redshift of z = 0.5. The highest displacement is for those galaxies that undergo major mergers and the lowest is for those that undergo non-merger interactions. The rms scatter we measure in log(SFR) based on the left panel of Fig. 2 is 0.44 dex and the enhancement in SFR for the interacting systems compared to the isolated system in the mass bin of 10$^{7.0 - 8}$ M$\odot$ is 3.35.  As the redshift and nature of interaction in our sample are different, we cannot directly compare our observed results with those obtained from these latter simulations.  Our sample contains galaxies that are post-merger products, which undergo minor mergers and non-merger interactions and have a range of projected separation. However, it is still interesting to see that our observed values are comparable with the simulation results, suggesting a role of interactions in the enhancement of SFR in our sample. We note that the SFR used in the log(SFR) versus stellar-mass plot (Fig. 2) and in Fig. 8 of \cite{Martin2021} is the current or instantaneous SFR of the galaxies. 
\section{Summary}
The effect of interactions on the evolution of low-mass galaxies is not well understood. 
In the present study, we performed a UV study of a sample of 22 interacting and 36 single gas-rich dwarf galaxies in the Lynx-Cancer void region using FUV images from the GALEX mission. We estimated their instantaneous SFR from their FUV luminosity in order to understand the effect of interactions on their SFR.  We find an enhancement in SFR by a factor of 3.4$\pm$1.2 for the interacting systems compared to single dwarf galaxies in the stellar mass range 10$^{7}$ - 10$^{8}$ M$\odot$. This value is comparable to the enhancement found by previous observational studies in the SFR of low-mass interacting galaxies, with stellar masses of $\sim$ 10$^9$ M$\odot$. Also similar to the predictions based on the simulation of dwarf galaxies, in the mass range of 10$^{7.5 - 9}$ M$\odot$ at a redshift of $\sim$ 0.5. Our results suggest  that the dwarf--dwarf galaxy interactions can lead to an enhancement in their SFR. Although our sample contains fewer galaxies, this study provides the first quantitative insights into the nature of interactions of dwarf 
galaxies in the sub-10$^8$ M$\odot$ regime 
and increases the small number of interacting dwarfs in the local Universe studied in the FUV. In future, we plan to study a larger sample of dwarf galaxies in different stages of interactions (major and minor mergers, fly-bys, etc.) in order to understand the effect of these different stages in their SFR. Future deeper and higher-spatial-resolution UV studies will help to improve our understanding of the effect of dwarf-galaxy interactions on the spatial distribution of star forming clumps and will also help us to identify star formation in tidal tails.

\section*{Acknowledgements}
 We thank the referee for the insightful suggestions which have improved the manuscript significantly. SS acknowledges support from the Science and Engineering Research Board of India through Ramanujan Fellowship and POWER grant (SPG/2021/002672). This work used Astropy and Matplotlib software packages \citep{astropy:2013, astropy:2018, astropy:2022, Hunter:2007}.
\bibliography{biblio.bib}

\end{document}